\documentclass[a4paper,english,manuscript]{revtex4}
\usepackage[T1]{fontenc}
\usepackage[latin1]{inputenc}
\usepackage{amsmath}
\usepackage{amssymb}

\makeatletter
\@ifundefined{textcolor}{}
{%
 \definecolor{BLACK}{gray}{0}
 \definecolor{WHITE}{gray}{1}
 \definecolor{RED}{rgb}{1,0,0}
 \definecolor{GREEN}{rgb}{0,1,0}
 \definecolor{BLUE}{rgb}{0,0,1}
 \definecolor{CYAN}{cmyk}{1,0,0,0}
 \definecolor{MAGENTA}{cmyk}{0,1,0,0}
 \definecolor{YELLOW}{cmyk}{0,0,1,0}
 }

\makeatother

\makeatother

\makeatother

\makeatother

\makeatother

\makeatother

\makeatother

\makeatother

\usepackage{babel}
\addto\shorthandsspanish{\spanishdeactivate{~<>}}

\makeatother

\usepackage{babel}

\begin{document}

\title{Maxwell-Cattaneo's equation and the stability of fluctuations in the
relativistic fluid}

\author{D. Brun-Battistini$^{1}$, A. Sandoval-Villalbazo$^{1}$, A. L. Garcia-Perciante$^{2}$}

\address{$^{1}$Depto. de Fisica y Matematicas, Universidad Iberoamericana,
Prolongacion Paseo de la Reforma 880, Mexico D. F. 01219, Mexico.}

\address{$^{2}$Depto. de Matematicas Aplicadas y Sistemas, Universidad Autonoma
Metropolitana-Cuajimalpa, Artificios 40 Mexico D.F 01120, Mexico.}
\begin{abstract}
Extended theories are widely used in the literature to describe the
relativistic fluid. The motivation for this is mostly due to the causality
issues allegedly present in the first order theories. However, the
decay of fluctuations in the system is also at stake when first order
theories \emph{that couple heat with acceleration} are used. In this
paper it is shown how the generic instabilities in relativistic fluids
are not present when a Maxwell-Cattaneo type law is introduced in
the system of hydrodynamic equations. Emphasis is made on the fact
that the stabilization is only due to the difference in characteristic
times for heat flux relaxation and instabilities onset. This gives
further evidence that Eckart's like constitutive equations are responsible
for the first order sytem exhibiting unphysical behavior.
\end{abstract}
\maketitle

\section{Introduction}

The transport equations for the relativistic fluid have been well
established within the framework of irreversible thermodynamics \cite{ck,degroor}.
However, it was shown that Eckart's constitutive equation, as a closure
equation for the heat flux including an acceleration term as driving
force \cite{Eckart1-1}, renders the system unstable \cite{HL} and
seems to allow propagation of signals at speeds larger than the speed
of light \cite{israel1}. This instability and acausality in the relativistic
gas lead to the formulation and use of higher order or extended theories
\cite{Israel2,Israel3,Zimdahl,Jou4}.

The definite proof of the violation of Onsager's regression of fluctuations
hypothesis was firstly shown by Hiscock and Lindblom eventhough they
interpreted this contradiction with the tenets of irreversible thermodynamics
as an early onset of instabilities \cite{HL}. In their paper it is
shown that fluctuations in the linearized system of equations grow
exponentially with a very small characteristic time. Meanwhile, buried
deep in the fundamental hypothesis of non-equilibrium thermodynamics,
Onsager's regression assumption assert that spontaneous fluctuations
of microscopic origin should decay following the linearized equations
for the state variables. It is this assumption that is violated when
Eckart's equation for the heat flux is used, as it is shown in Ref.
\cite{GRG09}.

As a result, Hiscock and Lindblom suggested in their work that, due
to the unphysical behavior predicted by Eckart's formalism, {}``standard''
first order theories should be discarded in favor of the second order
one developed by Israel. Even though the causality in the extended
theories was at the time verified, to the authors' knowledge the stability
within such framework was not assessed.

In this work we revisit the stability analisis for fluctuations within
the linearized system by using a Maxwell-Cattaneo type constitutive
equation for heat. This calculation not only validates the suggestion
in Ref. \cite{HL} but also sheds light on the {}``stabilization''
mechanism by showing that the dynamics of the fluctuations are dominated
by two competing drives: Eckart's constitutive equation and Cattaneo's
damping.

The rest of this paper is divided as follows. In Sect. II the system
of transport equations is established and the Maxwell-Cattaneo law
for the heat flux is introduced. The equations are linearized for
fluctuations around equilibrium values for the state variables in
Sect. III. The dispersion relation and its cualitative analysis are
shown in Sect. IV, while Sect. V includes the corresponding discussion
and final remarks.

\section{Relativistic Transport Equations}

The transport equations for a simple relativistic fluid are obtained
from the conservation equations for the particle and momentum-energy
fluxes, that is\begin{equation}
N_{;\mu}^{\mu}=0\label{eq:1}\end{equation}
and \begin{equation}
T_{;\nu}^{\mu\nu}=0\label{eq:2}\end{equation}
where $N^{\mu}$ is the particle four flux given by\begin{equation}
N^{\mu}=nu^{\mu}\label{eq:3}\end{equation}
and \begin{equation}
T_{\nu}^{\mu}=\frac{n\varepsilon}{c^{2}}u^{\mu}u_{\nu}+ph_{\nu}^{\mu}+\Pi_{\nu}^{\mu}+\frac{1}{c^{2}}q^{\mu}u_{\nu}+\frac{1}{c^{2}}u^{\mu}q_{\nu}\label{eq:4}\end{equation}
is the stress-energy tensor. Here $n$ is the particle number density,
$u^{\mu}$ the hydrodynamic four-velocity, $c$ the speed of light,
$\varepsilon$ the internal energy density per particle and $p$ the
hydrostatic pressure. The dissipative fluxes are the heat flux $q^{\nu}$
and the Navier tensor $\Pi_{\nu}^{\mu}$. The spatial projector $h_{\nu}^{\mu}$
for a $+++-$ signature is given by\begin{equation}
h_{\nu}^{\mu}=\delta_{\nu}^{\mu}+\frac{u^{\mu}u_{\nu}}{c^{2}}\label{eq:5}\end{equation}
which satisfies $h_{\nu}^{\mu}u^{\nu}=0$ for $\mu=1,...,4$, and
the orthogonality conditions implied in this 3+1 representation are\begin{equation}
u_{\mu}\Pi_{\nu}^{\mu}=0,\qquad q_{\mu}u^{\mu}=0\label{eq:6}\end{equation}
The introduction of Eq. (\ref{eq:3}) in Eq. (\ref{eq:1}) yields
the relativistic continuity equation, i. e. \begin{equation}
\dot{n}+n\theta=0\label{eq:7}\end{equation}
with $\theta=u_{;\nu}^{\nu}$. Both energy and momentum balances are
extracted from Eq. (\ref{eq:2}). For $\nu=1,2,3$ one obtains the
momentum balance\begin{equation}
\left(\frac{n\varepsilon}{c^{2}}+\frac{p}{c^{2}}\right)\dot{u}_{\nu}+\left(\frac{n\dot{\varepsilon}}{c^{2}}+\frac{p}{c^{2}}\theta\right)u_{\nu}+p_{,\mu}h_{\nu}^{\mu}+\Pi_{\nu;\mu}^{\mu}+\frac{1}{c^{2}}\left(q_{;\mu}^{\mu}u_{\nu}+q^{\mu}u_{\nu;\mu}+\theta q_{\nu}+u^{\mu}q_{\nu;\mu}\right)=0\label{eq:8}\end{equation}
Considering $\nu=4$ leads to a total energy balance from which the
mechanical energy needs to be substracted in order to establish the
heat flux. A shortcut leading directly to the internal energy equation
is to consider the projection $u^{\nu}T_{\nu;\mu}^{\mu}=0$ which
leads to

\begin{equation}
n\dot{\varepsilon}+p\theta+u_{,\mu}^{\nu}\Pi_{\nu}^{\mu}+q_{;\mu}^{\mu}+\frac{1}{c^{2}}\dot{u}^{\nu}q_{\nu}=0\label{9}\end{equation}
Using the relation $\dot{\varepsilon}=\left(\frac{\partial\varepsilon}{\partial n}\right)_{T}\dot{n}+\left(\frac{\partial\varepsilon}{\partial T}\right)_{n}\dot{T}$
together with the ideal gas law $p=nkT$ , which holds in the relativistic
case \cite{ck}, one readily obtains an evolution equation for $T$,
i. e.

\begin{equation}
nC_{n}\dot{T}+p\theta+u_{;\mu}^{\nu}\Pi_{\nu}^{\mu}+q_{;\mu}^{\mu}+\frac{1}{c^{2}}\dot{u}^{\nu}q_{\nu}=0\label{10}\end{equation}
The set of hydrodynamic equations for the relativistic fluid is thus
given by Eqs. (\ref{eq:7}), (\ref{eq:8}) and (\ref{10}). The set
is not complete and constitutive equations have to be introduced in
order to express the dissipative fluxes in terms of the state variables.
In 1940, Eckart proposed, from a purely phenomenological approach
by enforcing consistency with the second law of thermodynamics, a
coupling of heat with hydrodynamic acceleration additional to the
Fourier term \cite{Eckart1-1}. The resulting system of equations
was shown to yield unphysical behavior by Hiscock and Lindblom \cite{HL}.
Following their suggestion, we here study the behavior of the linearized
system by introducing for the heat flux, a Maxwell-Cattaneo constitutive
relation. Such relation consists in an evolution equation for the
heat flux, that is

\begin{equation}
\tau\dot{q}^{\nu}+q^{\nu}=-\kappa h^{\mu\nu}\left(T_{,\nu}+\frac{T}{c^{2}}\dot{u}_{\nu}\right)\label{eq:11}\end{equation}
where $\tau$ is a relaxation time such as $\tau=D_{T}/C_{s}$, with
$C_{s}$ the speed of sound. Notice that Eq.(\ref{eq:11}) becomes
Eckart's constitutive equation \cite{Eckart1-1} if $\tau=0$.

\section{Linearized transport equations}

In this section a set of linearized equations will be obtained by
assuming small fluctuations of the state variables around their equilibrium
values (denoted by a subscript $0$), that is

\begin{equation}
F=F_{0}+\delta F\label{eq:12}\end{equation}
Once this substitution is made, second (and higher order) terms on
fluctuations are neglected. Then, for the continuity equation we have

\begin{equation}
\delta\dot{n}+n_{0}\delta\theta=0\label{eq:13}\end{equation}
and for the momentum balance

\begin{equation}
\frac{1}{c^{2}}\left(n_{0}\varepsilon_{0}+p_{0}\right)\delta\dot{u}_{\nu}+kT\delta n_{,\nu}+nk\delta T_{,\nu}-\zeta\delta\theta_{,\nu}-2\eta\left(\delta\tau_{;\nu}^{\mu}\right)_{;\mu}+\delta\dot{q}_{\nu}=0\label{eq:14}\end{equation}
where $\zeta$ and $\eta$ are the bulk and shear viscosity respectively;
while the heat equation in this linear approximation reads

\begin{equation}
nC_{n}\delta\dot{T}+n_{0}kT_{0}\delta\theta+\delta q_{;\nu}^{\nu}=0\label{eq:15}\end{equation}
Finally, Maxwell-Cattaneo's equation in terms of fluctuations is given
by

\begin{equation}
\tau\delta\dot{q}^{\nu}+\delta q^{\nu}=-\kappa h^{\mu\nu}\left(\delta T_{,\nu}+\frac{T_{0}}{c^{2}}\dot{\delta u}_{\nu}\right)\label{eq:16}\end{equation}
Notice that Eqs. (\ref{eq:13}) and (\ref{eq:15}) only depend on
the velocity through $\delta\theta$. Because of that, it is convenient
to calculate the divergence of Eq. (\ref{eq:14}) . Following this
standard procedure, the longitudinal velocity gradient fluctuation
mode is uncoupled and the momentum balance is reduced to a scalar
equation for $\delta\theta$\begin{equation}
\frac{1}{c^{2}}\left(n_{0}\varepsilon_{0}+p_{0}\right)\delta\dot{\theta}_{;\nu}+kT_{0}\nabla\delta n+n_{0}k\nabla^{2}\delta T-\left(\zeta+\frac{4}{3}\eta\right)\nabla^{2}\delta\theta+\delta q_{;\nu}^{\nu}=0\label{eq:16a}\end{equation}
Now both the internal energy and momentum linearized balance equations
depend on the divergence of the heat flux fluctuations. Thus, if one
calculates de divergence of the linearized Maxwell-Cattaneo equation,
Eq. (\ref{eq:16}), the system is reduced to a system of 4 scalar
equations for the unknowns $\text{\ensuremath{\delta}n}$, $\delta\theta$,
$\delta T$ and $\delta q_{;\nu}^{\nu}$. The cualitative analysis
of such system is carried out in the next section in the Fourier-Laplace
space in order to address the behavior of such fluctuations.

\section{The dispersion relation}

In order to establish the dispersion relation, the system matrix is
formed with Laplace and Fourier transforms of equations (\ref{eq:13}),
(\ref{eq:15}) and (\ref{eq:16a}) and the divergence of Eq. (\ref{eq:16}),
that is:

\begin{equation}
A=\left(\begin{array}{cccc}
s & n_{0} & 0 & 0\\
-kT_{0}q^{2} & n_{0}ms+q^{2}\left(\zeta+\frac{4\eta}{3}\right) & -n_{0}kq^{2} & \frac{s}{c^{2}}\\
0 & n_{0}kT_{0} & C_{n}n_{0}s & 1\\
0 & \frac{sT_{0}\kappa}{c^{2}} & -\kappa q^{2} & \tau s+1\end{array}\right),\label{eq:17}\end{equation}
with $s$ and $q$ the Laplace and Fourier parameters respectively.
The corresponding dispersion relation for the system can thus be written
as

\begin{equation}
d_{4}s^{4}+d_{3}s^{3}+d_{2}s^{2}+d_{1}s+d_{0}=0\label{eq:18}\end{equation}
where the coefficients $d_{0},\, d_{1},\, d_{2}$ y $d_{4}$ are given
by:

\begin{equation}
d_{0}=\frac{kT_{0}\kappa}{C_{n}mn_{0}}q^{4}\label{eq:19}\end{equation}

\begin{equation}
d_{1}=\frac{kT_{0}}{m}\left[\left(1+\frac{k}{C_{n}}\right)q^{2}+\frac{1}{n_{0}}\left(\zeta+\frac{4}{3}\eta\right)\frac{\kappa}{C_{n}n_{0}}q^{4}\right],\label{eq:20}\end{equation}

\begin{equation}
d_{2}=q^{2}\left[\frac{1}{mn_{0}}\left(\zeta+\frac{4}{3}\eta\right)+\frac{\kappa}{C_{n}n_{0}}\left(1-\frac{2kT_{0}}{mc^{2}}\right)+\frac{kT_{0}}{m}\left(1+\frac{k}{C_{n}}\right)\tau\right],\label{eq:21}\end{equation}

\begin{equation}
d_{3}=1+\frac{1}{mn_{0}}\left(\zeta+\frac{4}{3}\eta\right)\tau q^{2},\label{eq:22}\end{equation}

\begin{equation}
d_{4}=\tau-\frac{T_{0}\kappa}{c^{4}mn_{0}}\label{eq:23}\end{equation}
In order to determine approximate solutions for the fourth order dispersion
relation, we follow the same ideas as in Ref. \cite{pre09}. Since
$d_{4}$ is a small quantity, three of the roots of Eq. (\ref{eq:18})
are calculated by neglecting the fourth order term and solving for\begin{equation}
d_{3}s^{3}+d_{2}s^{2}+d_{1}s+d_{0}=0\label{eq:23aa}\end{equation}
which, using Mountain's method \cite{Mountain}, can be shown to yield
a real root given by\begin{equation}
s_{1}=-\frac{D_{T}}{\gamma}\label{eq:23a}\end{equation}
and the two imaginary solutions \begin{equation}
s_{2,3}=-\frac{D_{V}}{2}-\frac{D_{T}}{5}-z\left(\frac{5}{6}c^{2}\tau-D_{T}\right)q^{2}\pm i\sqrt{\frac{5}{3}\frac{kT}{m}}q\label{eq:23b}\end{equation}
where use has been made of the fact that for an ideal gas, for $z<<1$,
$C_{n}=\frac{3}{2}k$, $\gamma=\frac{5}{3}$ and the definitions $D_{V}=\frac{1}{mn_{0}}\left(\zeta+\frac{4\eta}{3}\right)$
and $D_{T}=\frac{\kappa}{n_{0}C_{n}}$ for the viscous and thermal
diffusivities have been introduced. These roots give rise to the usual
behavior of density fluctuations, that is, a decaying mode with a
characteristic time given by $s_{1}$ and two oscillating ones with
frequencies given by the imaginary part of $s_{2}$ and $s_{3}$.
Notice that the only relativistic corrections to this behavior corresponds
to the Stokes-Kirchoff coefficient, the term in parenthesis in Eq.
(\ref{eq:23b}). Clearly, since the characteristic time $\tau$ is
of the order of the mean collision time $\sim10^{-11}s$ and typical
thermal diffusivities are in the $10^{-5}-10^{-7}m^{2}s^{-1}$ range,
the first term dominates and the relativistic correction yields only
a slight increase in the real part of the conjugate roots which vanishes
as $z\rightarrow0$ in the non-relativistic limit. It is also important
to point out the fact that the relativistic heat diffusion term, $zD_{T}$,
has opposite sign and, although small, would actually yield a small
growth in amplitude.

To approximate the fourth root, we use the information of the previous
solutions by assuming they are still approximate roots of the complete
equation, which is plausible given the smallness of $d_{4}$. Following
such proposal, we use the fact that the sum of all roots in an $n$-th
order polynomial is equal to the ratio of the coefficients of the
$n-1$ and the $n$-th power to obtain\begin{align}
s_{4} & =-\left(\tau-\frac{3}{2}\frac{D_{T}z}{c^{2}}\right)^{-1}\left[1+\frac{q^{2}}{mn_{0}c^{4}}\left(D_{V}T_{0}\kappa+\frac{2}{3}\frac{T_{0}\kappa^{2}}{kn_{0}}-\frac{4}{3}\frac{T_{0}^{2}\kappa^{2}}{c^{2}mn_{0}}\right.\right.\label{eq:23c}\\
 & \left.\left.-\frac{2}{3}\frac{c^{4}m\kappa\tau}{k}+\frac{4}{3}c^{2}T_{0}\kappa\tau+\frac{5}{3}\frac{kT_{0}^{2}\kappa\tau}{m}+\frac{5}{3}c^{4}kn_{0}T_{0}\tau^{2}\right)\right]\nonumber \end{align}
Notice that, by taking $\tau=0$, the fourth root is positive and
thus corresponds to a growing mode. In Ref. \cite{pre09}, the existence
of this positive large solution was used to argue against Eckart's
constitutive equation, in particular the coupling of heat with acceleration.
For $\tau\neq0$, we can approximate \begin{equation}
s_{4}\sim-\frac{1}{\tau}+q^{2}\left(D_{T}+\frac{5}{3}\frac{kT_{0}}{m}\tau\right)\label{eq:25}\end{equation}
Now, Eq. (\ref{eq:25}) sheds more light on the issue since, by using
the fact that $\tau\gg\frac{T_{0}\kappa}{\rho c^{4}}$, one can conclude
not only that the system of hydrodynamic equations with Cattaneo's
closure (Eq. (\ref{eq:16})) is stable but also realize that this
stabilization takes place as the relaxation term dominates over the
exponential growth caused by Eckart's coupling. In other words, the
time derivative term in Maxwell-Cattaneo's constitutive equations
yields not only a modified time scale for the evolution of the fluctuations
but, more importantly, changes the overall physical behavior because
of the different signs of the terms in the denominator of Eq. (\ref{eq:25}).
The root found in Ref. \cite{pre09}\begin{equation}
s_{4}\sim\frac{c^{4}mn_{0}}{\kappa T_{0}}\label{eq:26}\end{equation}
yields an exponential growth in the structure factor while the result
in the current calculation\begin{equation}
s_{4}\sim-\frac{1}{\tau}\label{eq:27}\end{equation}
corresponds to a finite spectrum which could eventually be observed
and measured.

\section{Summary and final remarks }

In the previous section, a dispersion relation for the system of linearized
hydrodynamic equations for the relativistic fluid was obtained with
a Maxwell-Cattaneo heat evolution equation. Such procedure produced
a fourth order equation for the Fourier variable $s$ for a given
wavenumber $q$. This result is similar to the one obtained in Ref.
\cite{pre09} but only in structure since, as was emphatically pointed
out in Sect. IV, the overall behavior of the system changes radically.
This can be clearly seen by comparing the approximate roots in Eqs.
(\ref{eq:26}) and (\ref{eq:27}). The former yields an exponential
growth in the corresponding structure factor which destroys, theoretically,
the spectrum which is clearly unacceptable. However, the latter predicts
a decay in time of the fluctuations with a characteristic time determined
by $\tau$, the relaxation parameter introduced by Cattaneo.

In this way, the stabilization of the system predicted intuitively
by Hiscock and Lindblom in 1985 is confirmed. To our knowledge, only
the causality issue had been addressed so far. And moreover, the mechanism
for the stabilization can be qualitatively attributed to the competing
terms in the denominator of $s_{4}$ as given by Eq. (\ref{eq:25}).
By confirming that the use of a relaxation equation for the heat flux
eliminates the instability in the system of equation by dominating
over the runaway solution that is introduced by the acceleration-heat
coupling, verifies the fact that it is such relation the one which
in the first place led to the unphysical behavior of thermodynamic
fluctuations in the relativistic gas. We thus conclude that the stability
of the fluid within Maxwell-Cattaneo's formalism is in complete agreement
with the claim that heat cannot be coupled with acceleration.

\section*{Acknowledgements}

The authors wish to thank L. S. Garcia-Colin and H. Mondragon-Suarez
for their valuable comments for this work.

\end{document}